# On-chip Erbium-doped lithium niobate microcavity laser

Yi'an Liu[1,*], Xiongshuo Yan[1,*], Jiangwei Wu[1], Bing Zhu[1],

Yuping Chen[1,*] and Xianfeng Chen[1,*]

[1]*State Key Laboratory of Advanced Optical Communication Systems and Networks,
School of Physics and Astronomy, Shanghai Jiao Tong University, Shanghai 200240, China*
﹡These authors contributed equally to this Letter.
Corresponding authors : ※ypchen@sjtu.edu.cn; xfchen@sjtu.edu.cn

**Abstract:** The commercialization of lithium niobate on insulator (LNOI) wafer has sparked significant on-chip photonic integration application due to its remarkable photonic, photoacoustic, electro-optic and piezoelectric nature. A variety of on-chip LNOI-based optical devices with high performance has been realized in recent years. Here we developed 1 mol% erbium-doped LN crystal and its LNOI wafer, and fabricated an erbium-doped LNOI microdisk with high quality (∼ $1.05 \times 10^5$). C-band laser emission with ∼1530 nm and ∼1560 nm from the high-Q erbium-doped LNOI microdisk was demonstrated both with 974 nm and 1460 nm pumping, and the latter has better thermal stability. This microlaser would play an important role in the photonic integrated circuits of lithium niobate platform.

## 1 Introduction

Lithium niobate on insulator (LNOI) has been a research hotspot in recent years. Due to the excellent physical properties of lithium niobate, such as nonlinearity, electro-optic, acousto-optic, and piezoelectric effects, many on-chip devices have been developed, such as frequency doubler, modulator and filter, etc [1-8]. In order to realize the complete photonic integrated circuits (PIC) on lithium niobate (LN) chips and meet the requirements of future high-speed optical communications, the C-band light source on LN chips needs to be developed urgently. It is known that rare earth erbium ion doping can produce lasers in the C-band, and many kinds of $Er^{3+}$-doped lasers based on waveguides are developed [9,10]. However, these $Er^{3+}$-doped lasers are difficult to be integrated on-chip. Considering that the on-chip laser requires a small footprint for PIC, the high-quality-factor whispering gallery mode (WGM) microdisk or microring resonator laser would be a good choice [11-17]. Therefore, the erbium-doped microdisk laser on LN chip may benefit the research for PIC. It is noted that $Er^{3+}$ doped in the LNOI by ion implantation has been achieved in recent research, but due to the low doping concentration and nonuniform distribution of $Er^{3+}$, the device does not show good laser output characteristics [18]. Here, we directly grow the erbium-doped LN crystal and make it into LNOI wafer. Then, a 150-μm-diameter microdisk cavity was fabricated on $Er^{3+}$-doped 600-nm-thick z-cut LNOI and the C-band laser output from the microdisk was observed. The laser threshold of 974 nm and 1460 nm pump are 2.99 mW and 14.18 mW, respectively. The LNOI light source realizes a big step forward in the development of LN PIC.

## 2  Fabrication

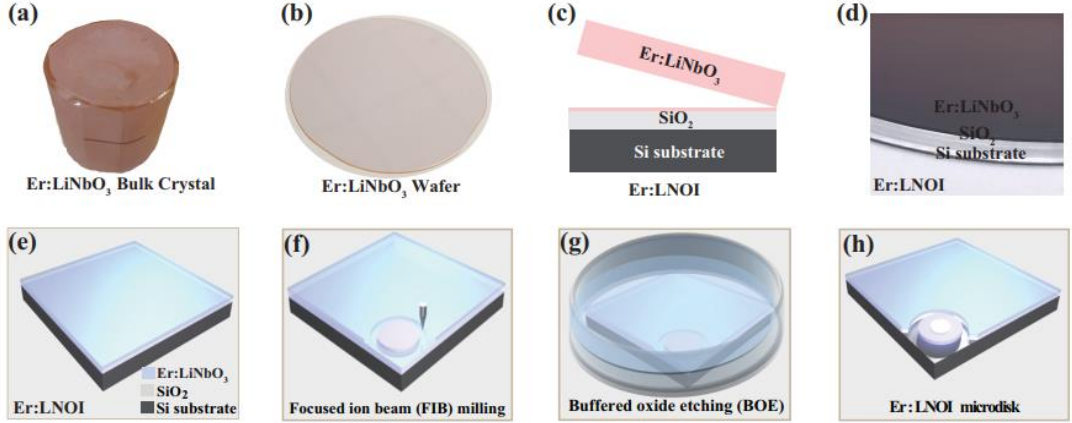

**Figure 1.** (a) - (d), the fabrication process of Er$^{3+}$-doped LNOI, where (a), (b) and (d) are pictures of real samples. (e) - (h), the fabrication process of the microdisk.

1% mol Er$^{3+}$ ion is doped into lithium niobate (LN) in the crystal growth process, shown in Fig. 1 (a). The Er$^{3+}$-doped 600-nm-thick z-cut LNOI, with 2 μm silica and 400 μm silicon substrate, is fabricated by smart-cut method from a 3 inch wafer [19], shown in Fig. 1 (b) - (d). The microdisk with 150-μm-diameter was fabricated by the focused ion beam (FIB) milling (ZEISS Auriga), shown in Fig. 1(f). It is worth stating, after FIB milling, a microring scanning pattern was used to remove residual in the edge of the microdisk, which is beneficial to get a smooth sidewall and achieve high optical Q factor. Then, the sample was immersed in buffered oxide etching (BOE) solution to form the silica pedestal under the Er$^{3+}$-doped LNOI microdisk, shown in Fig. 1(g) and (h). Usually, the silica pedestal has a nearly round border by BOE due to the isotropic corrosion of silica. However, in our device, we found that the silica pedestal is not a regular circle. The reason should be the silicon dioxide of the intermediate layer is not uniformly bonded to the Er$^{3+}$-doped LN. It is worthy of stating that further steps, such as chemo-mechanical polishing can help achieving higher Q factor [20-22] due to the smoother surface morphology.

## 3  Results and discussion

Fig. 2(a) shows the experimental setup. Firstly, the Q-factor of the Er$^{3+}$-doped LNOI microdisk was characterized by a C-band tunable continuous-wave laser (New Focus TLB-6712, linewidth < 200 kHz, 1520-1570 nm). The polarization controller (PC) was used to control the polarization of the input light. The tapered fiber, made by the heating and pulling method, was used to couple light into and out of the Er$^{3+}$- doped LNOI microdisk. The waist diameter of tapered fiber is approximately ~1 μm.

The Er$^{3+}$-doped LNOI microdisk was placed on a precise 3D nano stage which is under the optical microscope. Therefore, we can adjust the contact point and distance between the tapered fiber and the Er$^{3+}$-doped LNOI microdisk flexibly, to optimize the coupling efficiency or controlling the lasing wavelength [11]. The transmitted light from the tapered fiber was then linked to an InGaAs photodetector (PD) and an oscilloscope (OSC) to monitor the transmission spectrum.

We scanned the C-band laser wavelength from 1550 nm to 1560 nm loaded with a relatively low optical power, which is enough to measure the transmission spectrum of the microdisk, to avoid the interference from thermal effects. The transmission spectrum of the microdisk was shown in Fig. 2(b). And the Q factor of the mode indicated by a dotted black frame in Fig. 2(b) was measured to be 1.05 ×10$^5$, shown in Fig. 2(c), which is slightly lower than our previous microdisk resonator [23] due to the absorption of Er$^{3+}$-doped. Then, we used a 974 nm and a 1460 nm LD light source (Golight Co., Ltd) to achieve efficient laser emission in the Er$^{3+}$-doped LNOI microdisk, depicted in Fig. 2(a). The strong green emission was observed by a CCD camera shown in the insets of Fig. 3 (a). The spectrum was measured by an optical spectrum analyzer (200 nm-1100 nm) under 974 nm pump, shown in Fig. 3(a). It should be noted the spectrum under 1460 nm pump is similar as the 974 nm pump. The green and red emission are produced by cooperation up-conversion (CUC) and excited-state absorption (ESA) of 974 nm pump characterized as Fig. 3(b) (i). And Fig. 3(b) (ii) shows the second-order CUC of 1460 nm pump generates green and red emission [24].

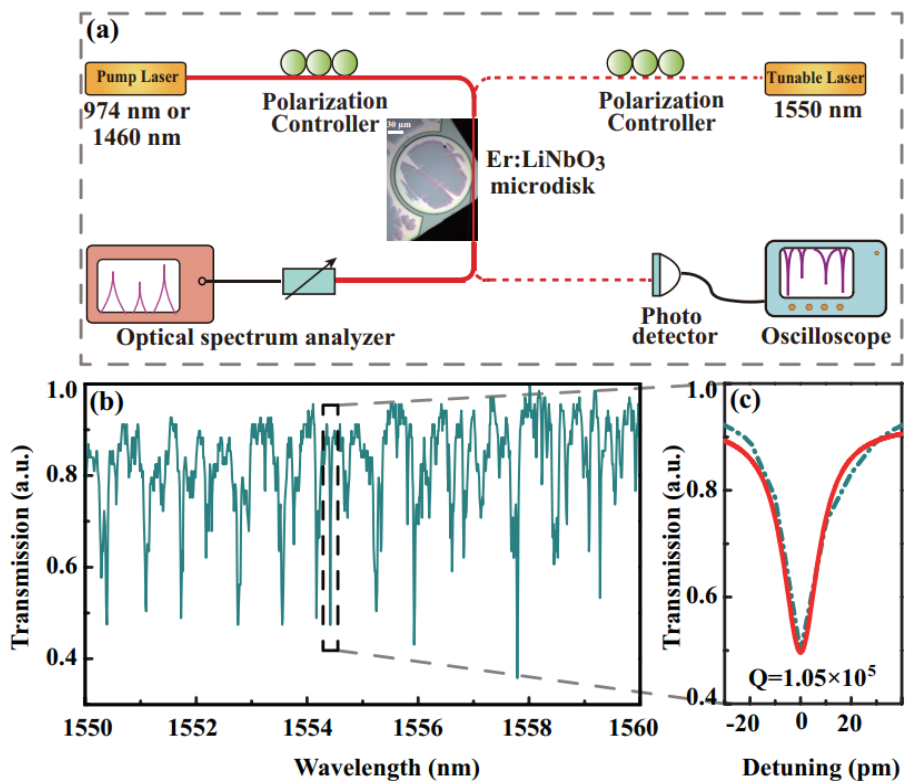

**Figure 2.** (a) Schematic of the experimental setup. (b) The transmission spectrum of the 150-μm-diameter Z-cut Er$^{3+}$-doped LNOI microdisk from 1550 nm to 1560 nm. (c) Lorentzian fitting of a measured mode around 1554.42 nm indicated by a dotted black frame in (b), exhibiting a Q factor of 1.05 ×10$^5$.

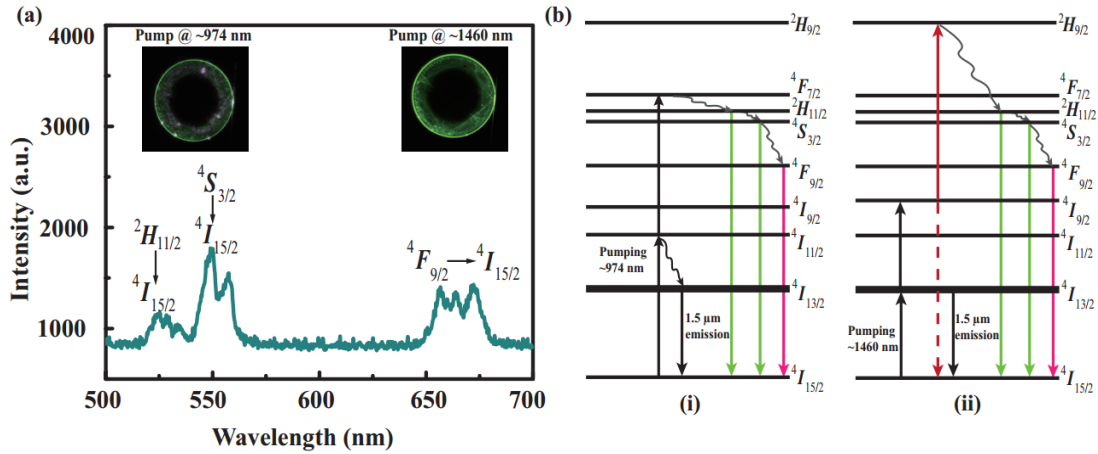

**Figure 3.** (a) Visible emission spectrum at 974 nm pumping of the $Er^{3+}$-doped LNOI microdisk. The two insets are optical microscope images of the visible emission from $Er^{3+}$-doped LNOI microdisk, with 974 nm and 1460 nm laser pumping, respectively. (b) Energy levels of $Er^{3+}$ involved in up-conversion emission. (i) The first-order CUC and ESA. (ii) The second-order CUC.

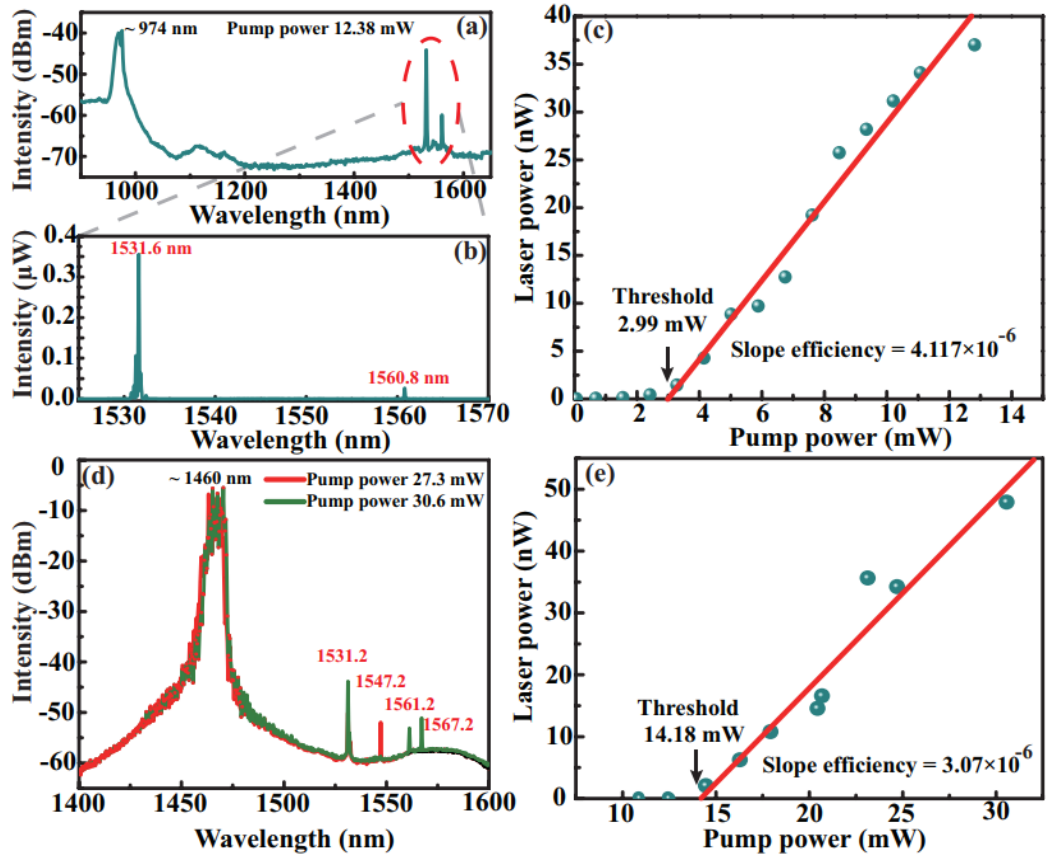

**Figure 4.** The $Er^{3+}$-doped LNOI microdisk laser spectrum and threshold characteristics. (a) The observed laser spectrum with 974 nm pump. (b) The magnified laser spectrum around 1531.8 nm and 1560.8 nm, shown in (a) by the red dotted ellipse. (c) The relationship between emitted laser power and 974 nm pump power. (d) The observed laser spectrum with 1460 nm pump at different pump power. (e) The relationship between emitted laser power and 1460 nm pump power.

The $Er^{3+}$-doped LNOI microdisk laser spectrum was measured with an OSA (from 0.6-1.75

μm). Fig. 4 (a) and (b) show the laser spectrum at C-band, pumped by 974 nm laser. The laser output power at 1531.6 nm was recorded versus the pump power. And the threshold and slope efficiency were calculated as 2.99 mW and $4.117\times10^{-6}$, respectively, shown in Fig. 4(c). The laser spectrum at C-band, pumped by 1460 nm laser was also measured, shown in Fig. 4(d), with the calculated threshold and slope efficiency of 1531.2 nm laser are 14.18 mW and $3.07\times10^{-6}$, respectively. It is worth noting that the laser threshold under 1460 nm pump is higher than under the 974 nm pump, this may be caused by the different coupling efficiency of the two pump laser. We may introduce the chaos in microdisk to improve the coupling efficiency of the 1460 nm pump [25].

When measuring the threshold, we founded the laser wavelength drift with the pump power changing. As pump at 974 nm, the single laser emission at around 1560 nm exhibited red-shift with the pump power increasing, shown in Fig. 5(a). The laser wavelength linearly drifted with the pump power at a slope of 47.9 pm/mW, shown in Fig. 5(b). The redshift should arise from the temperature increasing in the microdisk, which is related to the pump laser. And extra temperature control devices could reduce the red-shift.

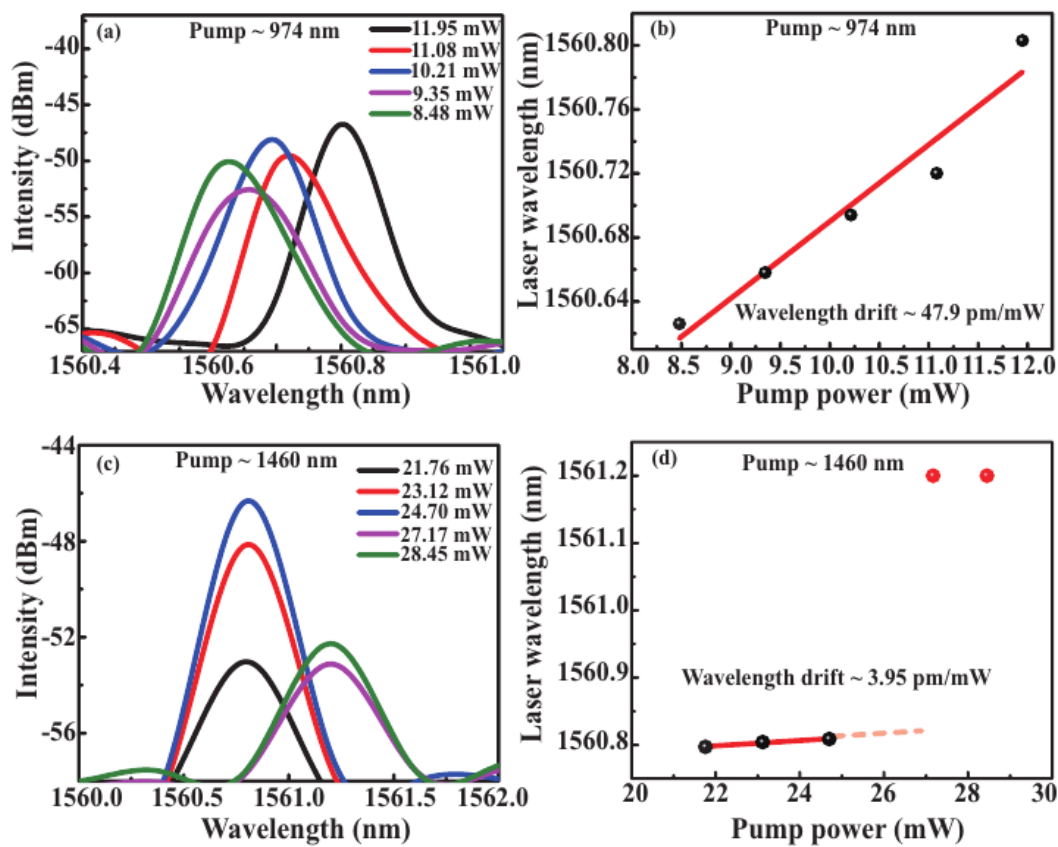

**Figure 5.** Laser wavelength drift with different pump power. (a) and (c), measured laser wavelength with different pump power. (b) and (d), the linear fit of laser wavelength drift.

With 1460 nm laser pumping, the laser emission wavelength at around 1561 nm drift was also observed, shown in Fig. 5(c). At a low level of the pump power (below 27 mW), the linearly drifted with the pump power was also calculated at a slope of 3.95 pm/mW, shown in Fig. 5(d). When the pump power higher than 27 mW, the laser wavelength abruptly changed about 0.4 nm. This could be explained by the mode competition in the inhomogeneous broadened lasers. Because of the tiny frequency distance between adjacent resonance modes in this relatively large multimode microdisk, the mode competition would appear with the pump power increasing, which leads to the laser wavelength changing [26]. Besides, by comparing the wavelength drift slope, we found pumping with 1460 nm laser has better thermal stability. The reason may be the 974 nm pump laser accompany with an ESA process which can store more pump laser energy in the microdisk. It is worth stating that more details about this phenomenon still need further study.

## 4 Conclusion

In summary, a high-Q $Er^{3+}$-doped LNOI microdisk was fabricated. Pumping with 974 nm and 1460 nm laser, both the visible emission and C-band laser emission were observed. The C-band laser threshold with 974 nm and 1460 nm pump are 2.99 mW and 14.18 mW, respectively. The laser wavelength drift with pump power of 974 nm and 1460 nm are also calculated and analyzed. Combining the properties of the dopant erbium with the excellent nonlinear, electro-optic, acousto-optic and integrated properties of the LNOI, the development of many class of new and higher functional devices could be expected in photonic integrated circuits, such as on-chip LN amplifier, tunable laser and so on.

**Acknowledgements**
We would like to acknowledge Shanghai Daheng Optics and Fine Mechanics Co., Ltd. and Jinan Jingzheng Electronics Co., Ltd. for the $Er^{3+}$-doped LNOI cooperative research and development. This work is supported by National Key R & D Program of China (Nos. 2019YFB2203500 and 2017YFA0303700), the National Natural Science Foundation of China (NSFC) (Nos. 91950107), Foundation for Development of Science and Technology of Shanghai (17JC1400400).

**Conflict of interest**
The authors declare that they have no conflict of interest.

## 5 References
[1] Wang C, Zhang M, Chen X, et al. Integrated lithium niobate electrooptic modulators operating at CMOS-compatible voltages[J]. Nature, 2018, 562(7725): 101-104.
[2] Li M, Ling J, He Y, et al. Lithium niobate photonic-crystal electro-optic modulator[J]. Nature Communications, 2020, 11(1): 1-8.
[3] Shao L, Yu M, Maity S, et al. Microwave-to-optical conversion using lithium niobate thin-film acoustic resonators[J]. Optica, 2019, 6(12): 1498-1505.
[4] Ge L, Chen Y, Jiang H, et al. Broadband quasi-phase matching in a MgO: PPLN thin film[J]. Photonics Research, 2018, 6(10): 954-958.
[5] Lin J, Yao N, Hao Z, et al. Broadband quasi-phase-matched harmonic generation in an on-chip monocrystalline lithium niobate microdisk resonator[J]. Physical review letters, 2019, 122(17): 173903.
[6] Boes A, Corcoran B, Chang L, et al. Status and potential of lithium niobate on insulator (LNOI) for photonic integrated circuits[J]. Laser & Photonics Reviews, 2018, 12(4): 1700256.

[7] Fang Z, Haque S, Lin J, et al. Real-time electrical tuning of an optical spring on a monolithically integrated ultrahigh Q lithium nibote microresonator[J]. Optics letters, 2019, 44(5): 1214-1217.

[8] Jiang H, Yan X, Liang H, et al. High harmonic optomechanical oscillations in the lithium niobate photonic crystal nanocavity[J]. Applied Physics Letters, 2020, 117(8): 081102.

[9] Brske D, Suntsov S, Rter C E, et al. Efficient ridge waveguide amplifiers and lasers in Er-doped lithium niobate by optical grade dicing and three-side Er and Ti in-diffusion[J]. Optics Express, 2017, 25(23): 29374-29379.

[10] Sohler W, Das B K, Dey D, et al. Erbium-doped lithium niobate waveguide lasers[J]. IEICE transactions on electronics, 2005, 88(5): 990-997.

[11] Min B, Kippenberg T J, Yang L, et al. Erbium-implanted high-Q silica toroidal microcavity laser on a silicon chip[J]. Physical Review A, 2004, 70(3): 033803.

[12] Polman A, Min B, Kalkman J, et al. Ultralow-threshold erbiumimplanted toroidal microlaser on silicon[J]. Applied Physics Letters, 2004, 84(7): 1037-1039.

[13] Gundavarapu S, Brodnik G M, Puckett M, et al. Sub-hertz fundamental linewidth photonic integrated Brillouin laser[J]. Nature Photonics, 2019, 13(1): 60-67.

[14] Cao Q T, Liu R, Wang H, et al. Reconfigurable symmetry-broken laser in a symmetric microcavity[J]. Nature communications, 2020, 11(1): 1-7.

[15] Latawiec P, Venkataraman V, Burek M J, et al. On-chip diamond Raman laser[J]. Optica, 2015, 2(11): 924-928.

[16] Jiang X, Shao L, Zhang S X, et al. Chaos-assisted broadband momentum transformation in optical microresonators[J]. Science, 2017, 358(6361): 344-347.

[17] Zhu S, Shi L, Xiao B, et al. All-optical tunable microlaser based on an ultrahigh-Q erbium-doped hybrid microbottle cavity[J]. ACS Photonics, 2018, 5(9): 3794-3800.

[18] Wang S, Yang L, Cheng R, et al. Incorporation of erbium ions into thin-film lithium niobate integrated photonics[J]. Applied Physics Letters, 2020, 116(15): 151103.

[19] Poberaj G, Hu H, Sohler W, et al. Lithium niobate on insulator (LNOI) for micro photonic devices[J]. Laser & photonics reviews, 2012, 6(4): 488-503.

[20] Wu R, Wang M, Xu J, et al. Long low-loss-litium niobate on insulator waveguides with sub-nanometer surface roughness[J]. Nanomaterials, 2018, 8(11): 910.

[21] Zhang J, Fang Z, Lin J, et al. Fabrication of crystalline microresonators of high quality factors with a controllable wedge angle on lithium niobate on insulator[J]. Nanomaterials, 2019, 9(9): 1218.

[22] Wang M, Wu R, Lin J, et al. Chemo-mechanical polish lithography: A pathway to low loss large-scale photonic integration on lithium niobate on insulator[J]. Quantum Engineering, 2019, 1(1): e9.

[23] Ge L, Jiang H, Zhu B, et al. Quality improvement and mode evolution of high-Q lithium niobate micro-disk induced by light annealing[J]. Optical Materials Express, 2019, 9(4): 1632-1639.

[24] Ye R, Xu C, Wang X, et al. Room-temperature near-infrared upconversion lasing in single-crystal Er-Y chloride silicate nanowires[J]. Scientific reports, 2016, 6(1): 1-6.

[25] Jiang X, Shao L, Zhang S X, et al. Chaos-assisted broadband momentum transformation in optical microresonators[J]. Science, 2017, 358(6361): 344-347.

[26] Svelto O, Hanna D C. Principles of lasers[M]. New York: Springer, 2010.